\documentclass[12pt]{article}
\usepackage{graphicx}

\usepackage[english]{babel}
\usepackage{amsmath,amssymb}
\usepackage{graphicx}

\newcommand\pubnumber{NuPhys2016-Maneschg}
\newcommand\pubdate{\today}

\def\napoli{Max-Planck-Institut f\"ur Kernphysik\\
Saupfercheckweg 1, D-69117 Heidelberg, GERMANY}

\def\Title#1{\begin{center} {\Large #1 } \end{center}}
\def\Author#1{\begin{center}{ \sc #1} \end{center}}
\def\Address#1{\begin{center}{ \it #1} \end{center}}

\newcommand\pubblock{\rightline{\begin{tabular}{l} \pubnumber\\
         \pubdate  \end{tabular}}}
\newenvironment{Abstract}{\begin{quotation}  }{\end{quotation}}
\newenvironment{Presented}{\begin{quotation} \begin{center} 
             PRESENTED AT\end{center}\bigskip 
      \begin{center}\begin{large}}{\end{large}\end{center} \end{quotation}}

\def\beq{\begin{equation}}
\def\eeq#1{\label{#1}\end{equation}}
\def\eeqn{\end{equation}}

\def\beqa{\begin{eqnarray}}
\def\eeqa#1{\label{#1}\end{eqnarray}}
\def\eeqan{\end{eqnarray}}

\let\bar=\overbar

\def\Dslash{\not{\hbox{\kern-4pt $D$}}}
\def\dslash{\not{\hbox{\kern-2pt $\del$}}}

\def\msb{{\bar{\ssstyle M \kern -1pt S}}}

\begin{document}
\begin{titlepage}
\pubblock

\vfill
\Title{Present status of neutrinoless double beta decay searches}
\vfill
\Author{Werner Maneschg}
\Address{\napoli}
\vfill
\begin{Abstract}
Several new generation experiments searching for neutrinoless double beta decay ($0\nu\beta\beta$) have become operational over the last five years. This report summarizes the present status of the experimental search and discusses peculiarities, challenges and reached half-life limits/sensitivities in these experiments. So far, no evidence for $0\nu\beta\beta$ has been found. Starting from the current situation, the paper addresses the question whether an experiment alone will be able to proof unambiguously $0\nu\beta\beta$ decay and which would be the key-requirements to succeed in this.
\end{Abstract}
\vfill
\begin{Presented}
NuPhys2016, Prospects in Neutrino Physics\\
Barbican Centre, London, UK,  December 12--14, 2016
\end{Presented}
\vfill
\end{titlepage}
\def\thefootnote{\fnsymbol{footnote}}
\setcounter{footnote}{0}

\section{Introduction}
Many extensions of the Standard Model of Particle Physics predict that neutrinos are their own anti-particles (Majorana particles) \cite{bib:majorana-neutrino}. One consequence thereof might be lepton number violation (LNV). This again would allow for a rare process called neutrinoless double beta decay ($0\nu\beta\beta$), in which an atom with mass and charge number $(A,Z)$ decays via $(A,Z+2)+2e^{-}$. The observation of $0\nu\beta\beta$ decay would have far-reaching implications. On one side, the observable half-life, i.e. $T_{1/2}^{0\nu}$, is inversely proportional to the effective neutrino mass $\langle m_{\beta\beta}\rangle$ and thus allows to establish the mass hierarchy of neutrinos. It allows to test predictions not only of left-right symmetric models, but also those including right handed currents and heavy neutral leptons \cite{bib:bsm-test}. Further, by linking lepton number violation with CP violation (leptogenesis) \cite{bib:leptogenesis}, neutrinos could be responsible for the unsolved problem in cosmology that barionic matter prevails over antimatter.\\
Over the last decade, $0\nu\beta\beta$ search has seen a boost with new experiments becoming operational and several more in preparation. Some of them use the $\beta\beta$ isotopes $^{136}$Xe, $^{76}$Ge, $^{130}$Te or $^{82}$Se, other ones plan to deploy also $^{48}$Ca, $^{100}$Mo, $^{116}$Cd or $^{150}$Nd. Even though different detection techniques are considered, they all measure the sum of the energies of the outgoing electrons released in the $\beta\beta$ decays. So, the non-LNV and already observed $2\nu\beta\beta$ decay channel leads to a beta-like spectrum, while $0\nu\beta\beta$ decays would lead to a Gaussian peak at the $Q_{\beta\beta}$ value. The latter corresponds to the mass difference of the parent and daughter nuclide. By looking at the leading experiments in the field, the $T_{1/2}^{0\nu}$ half-life has to be beyond 5$\times$10$^{25}$\,yr, i.e. 15 orders of magnitude longer than the age of the universe. This explains the stringent requirements to the experimental programs in terms of energy resolution, detector efficiencies and background suppression capabilities.\\
This paper reviews the present status of $0\nu\beta\beta$ decay searches by presenting the new generation experiments, that are already operational or will start presumably data collection in the next year. Detector performance, background conditions, hardware upgrades and $T_{1/2}^{0\nu}$ / $\langle m_{\beta\beta}\rangle$  sensitivities, that have been achieved or are designed, are outlined. Based on this information, the draft concludes with the scenario that $0\nu\beta\beta$ decay does exist and highlights the pre-requirements needed for a single experiment to proof unambiguously the existence of $0\nu\beta\beta$ decay.

\section{Experimental search for $0\nu\beta\beta$ decay in 2016}

\paragraph{CUORE} The Cryogenic Underground Observatory for Rare Events (CUORE) makes use of a bolometric technique to search for 0$\nu\beta\beta$ decays. TeO${_2}$ crystals are cooled down to (10-15)\,mK using helium inside a multi-layer copper cryostat surrounded by ancient Roman lead. Energy depositions are absorbed by the crystals and registered by thermistors as a temperature increase. Similar to Ge semiconductors, TeO${_2}$ crystals have excellent energy resolutions of 0.2\% at the $Q_{\beta\beta}$ value and large total detector efficiencies of 78-87\%.\\
As a first step towards CUORE, one tower ('CUORE-0') containing 52\,TeO${_2}$ crystals with a total mass of 39 kg was assembled. Correspondingly, the fiducial $\beta\beta$ mass of $^{130}$Te was 11\,kg. The observed background at $Q_{\beta\beta}$ is 0.058\,cts/(keV$\cdot$kg$\cdot$yr) and consists mainly of surface $\alpha$-events. The achieved sensitivity in 2015 after 2\,yr of data collection is $T_{1/2}^{0\nu}$ $>$2.7$\times$10$^{24}$\,yr (90\% C.L.). Together with the data set of the prototype Cuoricino it amounts to $T_{1/2}^{0\nu}$ $>$4$\times$10$^{24}$\,yr (90\% C.L.) \cite{bib:CUORE-0_2014}.\\
In the meantime, the CUORE detector array consisting of 19 CUORE-0 like towers with a total mass of 988\,kg and a $\beta\beta$ mass of 206\,kg has been completed. After cooling down and commissioning, data acquisition is scheduled to start in the first quarter of 2017 \cite{bib:CUORE_DiDomizio}. The designed sensitivity after a 5\,yr run will be $T_{1/2}^{0\nu}>$9.5$\times$10$^{25}$\,yr (90\% C.L.), which translates into $\langle m_{\beta\beta}\rangle<$(0.05-0.13)\,eV \cite{bib:CUORE_2014}.

\paragraph{GERDA} The GERmanium Detector Array (GERDA) consists of high purity germanium detectors enriched in $^{76}$Ge at (86-88)\%. Similar to bolometers, Ge diodes have excellent energy resolutions of 0.2\% and the obtained detector efficiencies are again high, namely 62-66\%. The diodes are mounted on low mass copper holders and submersed into liquid argon (LAr) inside a 64\,m$^3$ cryostat. The LAr serves as coolant and passive shield against external radiation.\\ 
In Phase I of the experiment (2011-2013) mainly detectors from former 0$\nu\beta\beta$ decay experiments were deployed, which amounted to 17.7\,kg. Even though an unexpectedly strong concentration of the contaminant $^{42}$Ar was encountered, it was possible to achieve a background of 0.01\,cts/(keV$\cdot$kg$\cdot$yr). By performing a blind analysis on the 21.6\,kg$\cdot$yr dataset, GERDA achieved a half-life limit of $T_{1/2}^{0\nu}>$2.1$\times$10$^{25}$\,yr (90\% C.L.) \cite{bib:GERDA_2013} and thus $\langle m_{\beta\beta}\rangle<$(0.2-0.4)\,eV \cite{bib:GERDA_2013}.\\
A two-years lasting upgrade phase followed, in which 30 new detectors of 20\,kg were produced, characterized and deployed in addition to the old ones. The design of the new detectors was optimized allowing for an improved energy resolution and enhanced pulse shape performance. Two LAr scintillation light read-outs were also developed and installed for further background reduction.\\
Phase II started in December 2015. First data were released in mid 2016 and comprised 10\,kg$\cdot$yr. A background of 0.001\,cts/(keV$\cdot$kg$\cdot$yr) was achieved for the new detectors. A half-life limit of $T_{1/2}^{0\nu}$ of $>$5.4$\times$10$^{25}$\,yr (90\% C.L.) was deduced \cite{bib:GERDA_2015}. A second data release is planned for 2017.\\
GERDA plans to gain an exposure of 100\,kg$\cdot$yr under quasi background-free conditions allowing for a half-life sensitivity of $T_{1/2}^{0\nu}>$2$\times$10$^{26}$\,yr after $\sim$4\,yr of operation.

\paragraph{KamLAND-Zen} The KAmioka Liquid Scintillator Anti-Neutrino Detector (KamLAND) is a multi-purpose detector originally designed for solar, geo- and reactor-neutrino measurements. In recent years the physics program was extended to 
0$\nu\beta\beta$ decay search leading to the KamLAND-Zen experiment. Herein, a smaller inner balloon containing xenon-loaded scintillator was installed in the center of the larger spherical vessel filled with 1000\,tons of ultra-radiopure organic liquid scintillator (LS). The latter one acts in this case as a superb active background veto.\\ 
In Phase I (2011-2012) of the experiment, a mini-balloon with R$<$1.54\,m and xenon enriched in $^{136}$Xe at 91\% was used. The $\beta\beta$ mass was 320\,kg, but the total detection efficiency after all cuts only 25\% and the energy resolution at $Q_{\beta\beta}$ is 10\%. An unknown strong background peak appeared close to $Q_{\beta\beta}$, which was identified as $^{110m}$Ag, a fall-out product of the Fukushima reactor accident. Based on an exposure of 89.5\,kg$\cdot$yr, a half-life limit of $T_{1/2}^{0\nu}$ $>$1.9$\times$10$^{25}$\,yr (90\% C.L.) was obtained \cite{bib:KamLAND-Zen_2013}. After Xe extraction, several scintillator purification campaigns followed reducing successfully the $^{110m}$Ag content.\\
In the following Phase II (2013-2016) 504\,kg$\cdot$yr of data were collected. This led to a half-life limit of $>$9.2$\times$10$^{25}$\,yr (90\% C.L.). By combining both results, the $T_{1/2}^{0\nu}$ limit becomes $>$1.1$\times$10$^{26}$\,yr (90\% C.L.) and $\langle m_{\beta\beta}\rangle<$(0.06-0.16)\,eV \cite{bib:KamLAND-Zen_2016}. These are the best limits achieved in the field so far.\\
In 2016, the KamLAND-Zen collaboration prepared the upgrade for Phase 3, in which a sensitivity of $\langle m_{\beta\beta}\rangle<$0.04\,eV should be reached. So, 750\,kg of enriched Xe were dissolved in LS
and the nylon mini-balloon was replaced with a two-fold larger and cleaner one. Recently, a leak was discovered, such that the latter one has to be replaced as well. Delayed by this, data collection will start in mid 2017 \cite{bib:KamLAND-Zen_Shirai}.

\paragraph{EXO} The Enriched Xenon Observatory (EXO) experiment uses a pressurized time projection chamber (TPC) filled with liquid xenon (LXe) as source and detection medium. The xenon is enriched in $^{136}$Xe at 81\% and is cooled down with a high-purity heat transfer fluid inside a radiopure copper cryostat. The cylindrical TPC has two wire grids at both ends with different applied voltages that lead to a drift field. Behind the grids large area avalanche photodiodes are mounted. Both detection methods combined allow to read out simultaneously charge and scintillation light produced by ionisation, as well as to reconstruct the 3D-position of events and to suppress background via pulse shape analysis.\\
In Phase I (2011-2014) of EXO-200, 150\,kg of enriched LXe were deployed. Data were collected for a total of 
100\,kg$\cdot$yr and with a high detector efficiency of 85\%. A background of 0.0017\,cts/(keV$\cdot$kg$\cdot$yr) was achieved, however at a modest energy resolution of 3.7\%. (31$\pm$4) events within 2$\sigma$ around $Q_{\beta\beta}$ were observed, still compatible with the background model \cite{bib:exo-200_2014}. A half-life limit of $T_{1/2}^{0\nu}>$1.1$\times$10$^{25}$\,yr (90\% C.L.) was established, which converts into $\langle m_{\beta\beta}\rangle<$(0.19-0.45)\,eV.\\
In Phase II (started in the first quarter of 2016), the collaboration succeeded to improve the energy resolution to 3\% and to obtain lower background levels via an improved pulse shape technique and reduced Rn levels. The goal sensitivity after a 3\,yr run will be $T_{1/2}^{0\nu}>$5.7$\times$10$^{26}$\,yr (90\% C.L.), correspondingly $\langle m_{\beta\beta}\rangle<$0.09\,eV \cite{bib:exo-200_Yen}.

\paragraph{Other Demonstrators} The {\it Majorana Demonstrator} operates 30\,kg of $^{76}$Ge enriched detectors in two large vacuum  cryostats embedded within a passive shield of ultrapure copper, lead and neutron moderators. A first module became operational at the beginning of 2016, the second module has been commissioned and data collection started in the second half of 2016. The background is at the designed level of 0.01\,cts/(keV$\cdot$kg$\cdot$yr). Start of data blinding is planned for the near future \cite{bib:Majorana_Elliott}. The entire setup will reach a half-life limit of $T_{1/2}^{0\nu}$ $>$4$\times$10$^{25}$\,yr after 1\,yr of running \cite{bib:MJ-DEM_2012}.\\
The {\it SuperNEMO Demonstrator} is based on the design principles of the NEMO-3 tracking calorimeter. It will consist of $\beta\beta$ emitting foils  sandwiched between multiwire chambers for track reconstruction, and calorimeter walls for energy determination. This will allow for the unique opportunity to reconstruct the full kinematics of background and 0$\nu\beta\beta$-like events. In 2016, two calorimeter walls were installed, commissioning of the trackers is planned in the first quarter of 2017 \cite{bib_SuperNEMO_Waters}. 7\,kg of $^{82}$Se are planned to be used, but also other isotopes will be considered. If successful, 20 Demonstrator modules might be installed afterwards in the framework of SuperNEMO. The goal sensitivity after a 5\,yr operation would be $T_{1/2}^{0\nu}>$1$\times$10$^{26}$\,yr (90\% C.L.), which corresponds to $\langle m_{\beta\beta}\rangle<$(0.04-0.10)\,eV.\\
Other experiments and R\&D projects are in the pipeline and some of them will become operational within the next years. A discussion about them is omitted here, since it is covered in another NuPhys2016 talk by S. Di Domizio.

\section{Roadmap for an unambiguous $0\nu\beta\beta$ discovery}
In the past it was repeatedly stated, that there is no $\beta\beta$ isotope and technique that ideally fulfill all the following advantages: a high $Q_{\beta\beta}$ value, a convenient $G^{0\nu}$ factor, a high natural isotopic fraction, enrichment possibilities, low isotope costs, low efforts in detector production, background suppression capabilities, high energy resolutions, easy and fast handling/operation of detectors. Further it was argued, that at least two independent experiments using different isotopes and thus different $Q_{\beta\beta}$ values would be needed to proof unambiguously the $0\nu\beta\beta$ discovery achieved by a certain experiment. So, waiting for a second experiment would be mandatory.\\
The first argument is certainly true. Regarding the second one, however, the question is, whether a single experiment will be able to provide a stringent proof, when discovering a $\gamma$-line at $Q_{\beta\beta}$ of a given isotope. The present situation of $0\nu\beta\beta$ decay experiments helps in figuring out the requirements for an experiment to check with high confidence a potentially observed $0\nu\beta\beta$-decay. The following four-steps roadmap can be defined:
\begin{enumerate}
\item {\bf Increase experimental sensitivity} The experimental $T_{1/2}^{0\nu}$ sensitivity has to be in reach of the $0\nu\beta\beta$ decay half-life. Former and present experiments mostly excluded the degenerate neutrino mass scenario. In the case of the inverted mass hierarchy, the $\langle m_{\beta\beta}\rangle$ is in the range of (0.01-0.04)\,eV. $T_{1/2}^{0\nu}$ sensitivities of (10$^{26}$-10$^{28}$)\,yr will be needed for a full coverage of this parameter space. Based on presently used technologies, several feasibility studies for experiments aiming at this goal are ongoing (KamLAND2-Zen, LEGeND, nEXO).
\item {\bf $\gamma$-peak detection at $Q_{\beta\beta}$ value} By establishing the proper experimental sensitivity, data should show a $\gamma$-line at the expected $Q_{\beta\beta}$-value. Other radioactive nuclear transitions or processes induced e.g. by neutrons should be excluded via nuclear data bases and background simulations.
\item {\bf Excluding unknown $\gamma$-lines that are not from $0\nu\beta\beta$ decays} Unknown nuclear processes such as forbidden transitions of higher order might appear in the spectrum and mimic a $0\nu\beta\beta$-decay $\gamma$-line. Two options can be pursued to overcome this difficulty:
\begin{itemize}
\item {Measure the topology of energy depositions:} $\beta\beta$ decays are single-site (SS) energy depositions, since the outgoing electrons are stopped within tiny detector volumes of few mm$^3$. On the contrary, $\gamma$-rays of similar energy often undergo Compton scattering and deposit the energy in multiple sites (MS) of a detector, with distances up to tens of cm from each other. Detectors operated in dense arrays such in GERDA might be used to reject MS events by detecting coincident signals. Single detectors that can measure the time-resolved pulse structure of signals might be able to distinguish directly between SS and MS events. Indeed, this so-called pulse shape analysis is a well established technique already in use by the Majorana and GERDA collaborations, as well as the EXO collaboration.  
\item {Extract the daughter nuclide:} A simultaneous detection of a $0\nu\beta\beta$-like signal and a shortly after extraction of a stable $0\nu\beta\beta$-decay daughter nuclide, e.g. $^{136}$Ba from $^{136}$Xe or $^{76}$Se from $^{76}$Ge, would proof the $\beta\beta$ nature of the parent nuclide. In solid state detectors, an extraction is not possible, but in a fluid. Indeed, the EXO collaboration is testing different techniques to isolate $^{136}$Ba atoms from the $\beta\beta$ detector mass. Most of these attempts still face different challenges: reduction of extraction times, increase of extraction efficiencies and adaption laboratory to the large-scale experiment conditions. 
\end{itemize}
\item {\bf Separation of $2\nu\beta\beta$ from $0\nu\beta\beta$ events:} Even though an experiment might succeed to discriminate SS from MS events with a very high efficiency, one type of background will still survive: $2\nu\beta\beta$ events. These produce SS signals and thus are $0\nu\beta\beta$-like events. Herein, the tail of the 2$\nu\beta\beta$ spectrum increasingly overlap with the $0\nu\beta\beta$ peak, the worse the energy resolution and the smaller the $2\nu\beta\beta$ half-life $T_{1/2}^{2\nu}$ is. For the isotopes used in the current experiments, $T_{1/2}^{2\nu}$ varies between 10$^{18}$ and 10$^{21}$\,yr. The most convenient isotopes are $^{136}$Xe and $^{76}$Ge with $T_{1/2}^{2\nu}$ half-lives of 2.17$\times$10$^{21}$ and 1.9$\times$10$^{21}$\,yr correspondingly. For both leading experiments, KamLAND-Zen and GERDA, the corresponding $2\nu\beta\beta$ spectra and $0\nu\beta\beta$-peaks using the currently achieved sensitivities are shown in Figures 1 and 2. The blue/green line corresponds to the spectrum excluding/including the energy resolution of the detectors. In the case of KamLAND-Zen, the energy resolution is only 10\% and $2\nu\beta\beta$ events swap largely into the $0\nu\beta\beta$-peak region. Contrarily for GERDA the energy resolution is 0.2\% and the $2\nu\beta\beta$ tail is still far away from the $0\nu\beta\beta$-peak.\\
\begin{figure}[t]
 \hspace{.09\linewidth}
 \begin{minipage}[b]{.45\linewidth}
  \label{fig:specxekamlandzen}
  \includegraphics[scale=0.30]{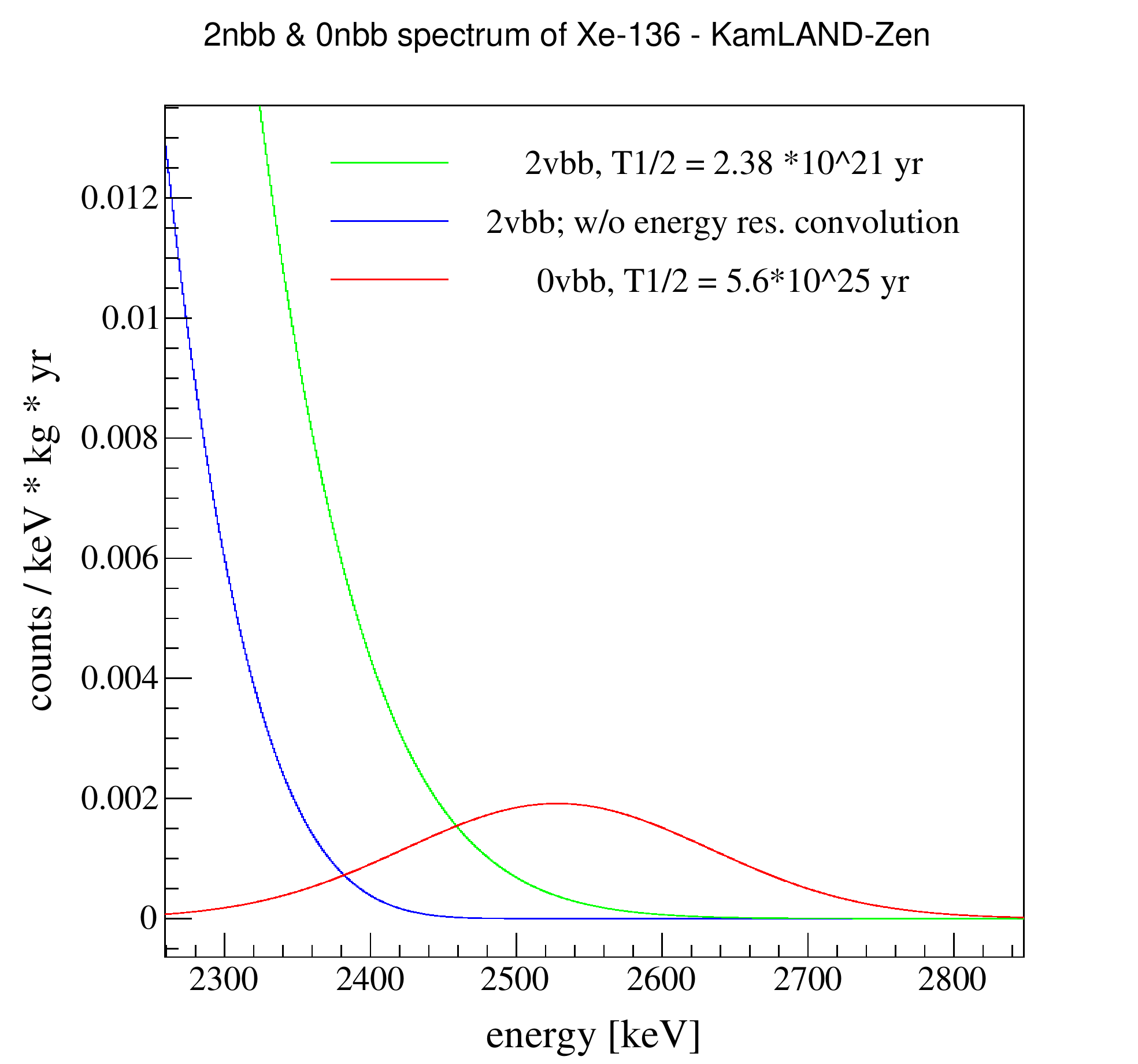}
  \caption{KamLAND-Zen: $2\nu\beta\beta$ vs. $0\nu\beta\beta$ for present $T_{1/2}^{0\nu}$($^{136}$Xe).} 
 \end{minipage}
 \hspace{.03\linewidth}
 \begin{minipage}[b]{.45\linewidth}
  \label{fig:specgegerda}
  \includegraphics[scale=.30]{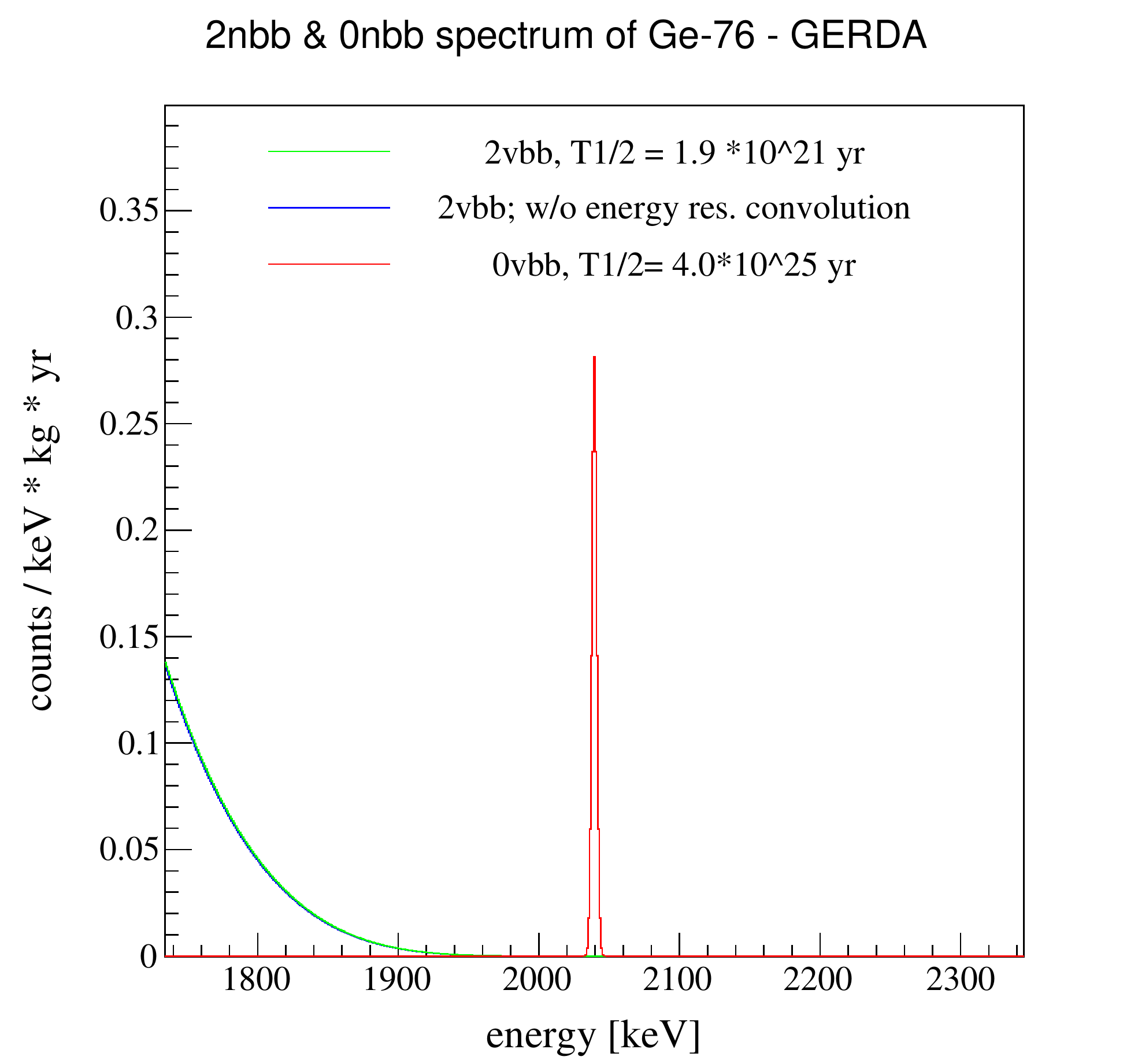}
  \caption{GERDA: $2\nu\beta\beta$ vs. $0\nu\beta\beta$ for present $T_{1/2}^{0\nu}$($^{76}$Ge).} 
 \end{minipage}
  \hspace{.05\linewidth}
\end{figure}
In order to assess the situation in the near future, it is opportune to have a look at the performance and projected sensitivity of current $\beta\beta$ experiments and define the following signal-to-background parametrisation: 
\begin{equation}
 R = 1 + B(2\nu\beta\beta)/S(0\nu\beta\beta)
\end{equation}
B($2\nu\beta\beta$) and S($0\nu\beta\beta$) stand for the $2\nu\beta\beta$ background and the 
$0\nu\beta\beta$ signal in the same energy window, respectively. Ideally, $R$ is equal 1. By applying a cut on $0\nu\beta\beta$ peak events one will loose $0\nu\beta\beta$ events, but also keep $R$ close to unity. An example is given in Figure 3 and 4: the cut is set here to 88\% $0\nu\beta\beta$ survival fraction.
For KamLAND-Zen and EXO, the $R$ parameter goes beyond 2 already for $T_{1/2}^{0\nu}$ half-lives of $\sim$1$\times$10$^{26}$ and $\sim$3$\times$10$^{26}$\,yr. For EXO it will be problematic after reaching 10$^{28}$\,yr. For GERDA, Majorana and CUORE it is not critical at all. Stronger cuts are not favorable, since only few $0\nu\beta\beta$ events are expected at high for $T_{1/2}^{0\nu}$ half-life beyond 10$^{26}$\,yr, even for large scale detectors with tonne scale $\beta\beta$ masses.
\begin{figure}
 \begin{minipage}[b]{.40\linewidth}
  \label{fig:specxekamlandzen}
  \includegraphics[scale=0.30]{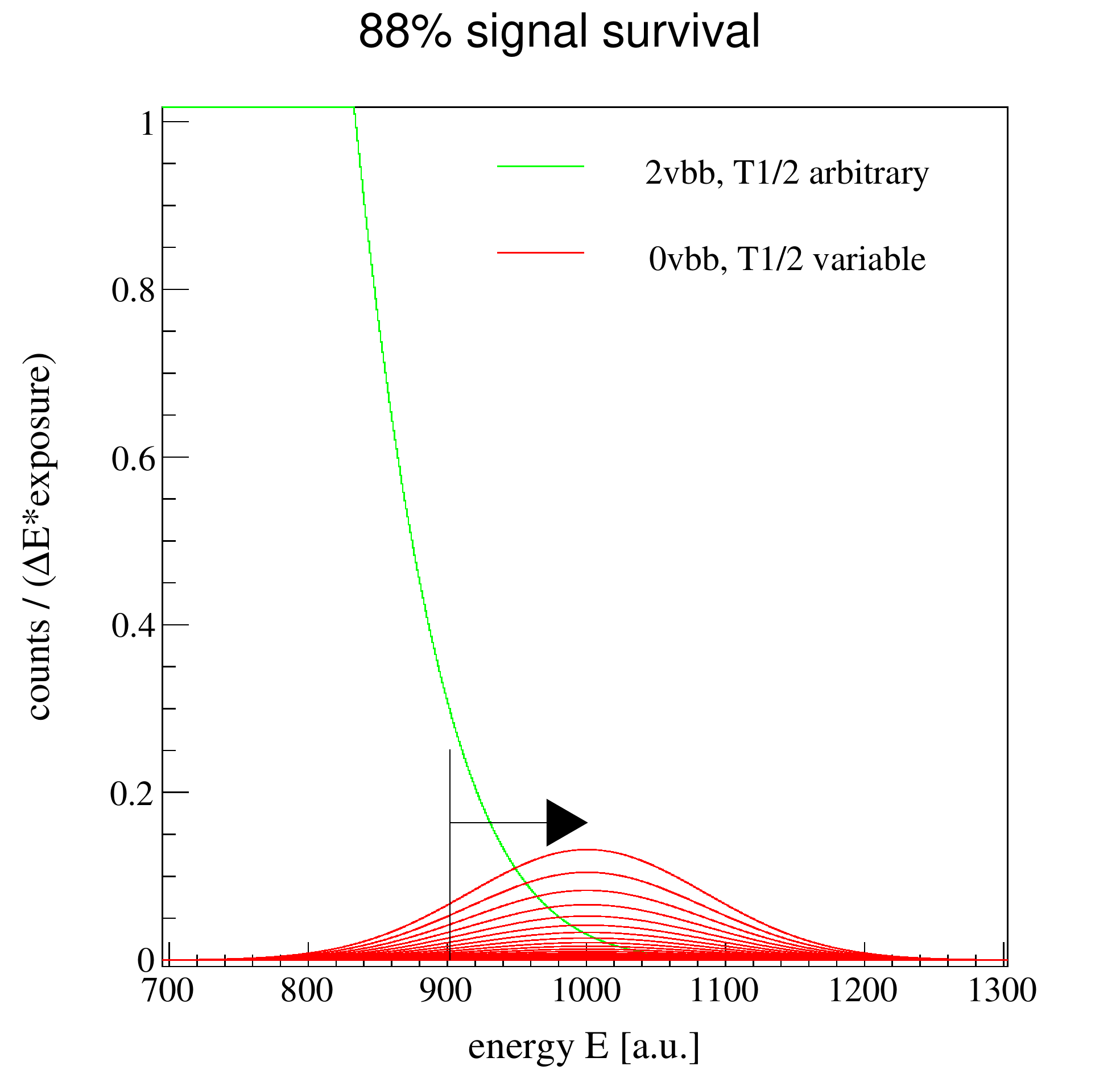}
  \caption{Sketch of a 88\% cut applied on $0\nu\beta\beta$ peak signals.} 
 \end{minipage}
 \hspace{.05\linewidth}
 \begin{minipage}[b]{.50\linewidth}
  \label{fig:specgegerda}
  \includegraphics[scale=.45]{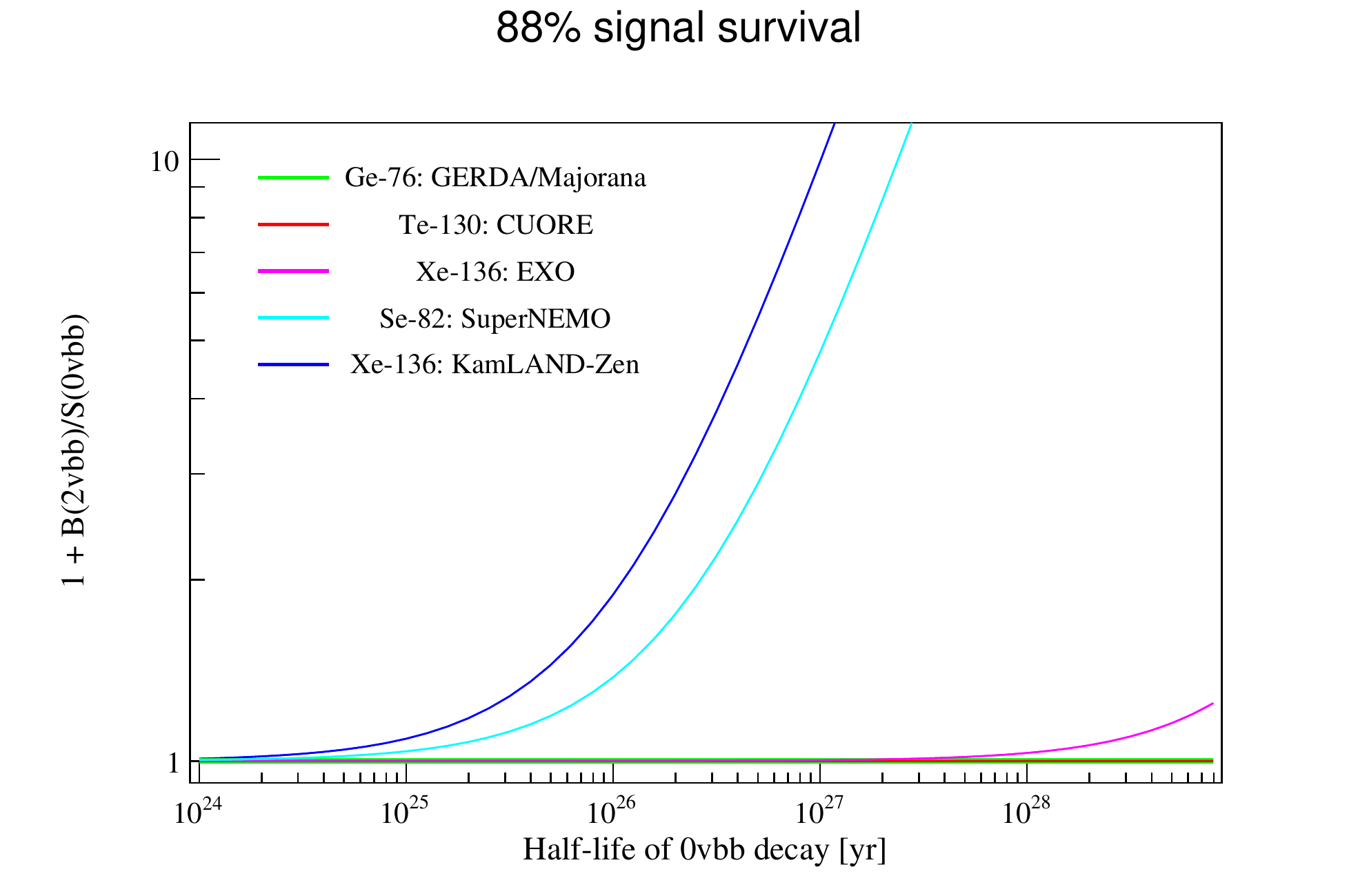}
  \caption{Parameter $R$ for projected sensitivities of current $\beta\beta$ experiments.} 
 \end{minipage}
\end{figure}
\end{enumerate}

\section{Conclusions} 
Several new generation $\beta\beta$ experiments have become operational in the last 5 years. Few of them have already concluded a first phase and - after substantial hardware upgrades - started/are preparing for next phases.\\
New generation experiments have reached half-life sensitivities of $T_{1/2}^{0\nu}\sim$10$^{25}$-10$^{26}$\,yr. The best sensitivities have been obtained by the KamLAND-Zen and GERDA experiment: 5.6*10$^{25}$ yr and 4.0*10$^{25}$ yr at 90\% C.L., correspondingly. KamLAND-Zen will probably be the first experiment being able to go beyond 10$^{26}$\,yr and to enter the inverted mass hierarchy regime.\\
Most experiments were affected by unexpected non negligible background components: KamLAND-Zen faced an $^{119m}$Ag pollution, GERDA experienced a high $^{42}$Ar concentration in the liquid argon, CUORE had to cope with surface $\alpha$ contamination and EXO with air-borne Rn. Nonetheless, in most cases it was possible to adopt counter actions and reach finally the designed background levels. Moreover, GERDA will be the first quasi background-free experiment for the goal sensitivity of $T_{1/2}^{0\nu}>$2$\times$10$^{26}$\,yr.\\
Beyond that, the experience from present $\beta\beta$ experiments teach us, that some experiments (i.e. large scale scintillator detectors) are more suitable for pushing the $T_{1/2}^{0\nu}$ sensitivity, while other approaches (i.e. solid state detectors) are more appropriate for a $0\nu\beta\beta$-peak discovery, under the assumption that excellent energy resolutions and quasi background-free conditions can be met.

\end{document}